\documentclass[twocolumn,reprint,aps,prl, floatfix,amsmath,amssymb,superscriptaddress]{revtex4-2}

\usepackage[utf8]{inputenc}  
\usepackage{graphicx}
\usepackage{bm}
\usepackage{braket}
\usepackage{xcolor} 
\usepackage{subfigure}     
\usepackage{diagbox}
\usepackage{amsthm}
\usepackage{hyperref}

\definecolor{dkgreen}{rgb}{0,0.6,0}

\definecolor{purple}{rgb}{0.5,0,0.5}

\begin{document}
	
	\title{A Study of Decay Rate of Bound Negative Muons}
	
	\author{Jian-Bo Deng}
	\email[Jian-Bo Deng: ]{dengjb@lzu.edu.cn}  
	\affiliation{Institute of Theoretical Physics $\&$ Research Center of Gravitation, Lanzhou University, Lanzhou, 730000, Gansu, China}
	\author{Miao-Yi Deng}
	\affiliation{Academy for Advanced Interdisciplinary Studies, Peking University, 100871, Beijing, China}
	\author{Shi-Jie Ma}\author{Rui-Bo Wang}\author{Qi-Qi Fan}\author{Peng-Zhang He}\author{Yi-Peng He}\author{Shuo-Wen Li}\author{Xian-Ru Hu}
	\affiliation{Institute of Theoretical Physics $\&$ Research Center of Gravitation, Lanzhou University, Lanzhou, 730000, Gansu, China}
	
	\date{\today}

	\begin{abstract}
	A number of experiments show that the decay lifetimes of muons bound to atomic nuclei are longer than the decay lifetimes of free muons. In this paper, a scheme of extending quantum mechanics (EQM) is proposed  to resolve this problem. The Schr$\ddot{\text{o}}$dinger's equation is obtained to prove the validation of this attempt. The decay ratio of bound muons is also calculated in EQM, and the result is in good agreement with the experimental data.
	\end{abstract}
	
	\maketitle

	\section{Introduction}
	\label{sec1}
	
	There have been a lot of experiments testing the decay rate of bound muons~\cite{p1,p2,p3,p4,p5,p6,p7,p8,p9,p10,p11,p12,p13,p14,p15}. A study that deserves attention is D. Apsel's work in ref.~\cite{p10}. In this research, the author summarized the experimental results of decay ratio of bound muons in numerous elements and tried to explain the experimental findings by studying the time delay effect in Bohr orbital theory. The ratio of decay rate takes
	\begin{equation}\label{eq2_1}
		R=\frac{1}{\gamma},
	\end{equation}
	where $\gamma=dt/d\tau$ is the time delay factor, $t$ is the particle's lifetime and $\tau$ is the particle's proper lifetime. In D. Apsel's theory~\cite{p16}, the proper time $d\tau$ of a particle moving in an electromagnetic field is suggested as
	\begin{equation}\label{eq2_2}
		cd\tau=\sqrt{-g_{\mu\nu}dx^{\mu}dx^{\nu}}+\frac{q}{mc}A_{\mu}dx^{\mu},
	\end{equation}
	where $g_{\mu\nu}$ is the metric of spacetime, $q$ is the electric charge of the particle, $m$ is the particle's rest mass, $c$ is the speed of light and $A_{\mu}$ is the electromagnetic four-potential. In D. Apsel's assumption, the bound muon moves in circular motion on the first Bohr radius orbit $r=r_{0}/Z$, where $Z$ is the nuclear charge number of the element, $r_{0}=4\pi\epsilon_{0}\hbar^{2}/(me^{2})$ is the Bohr radius, $\hbar$ is the reduced Planck constant, $\epsilon_{0}$ is the vacuum permittivity and $e$ is the elementary charge. Thus the total time dilation provides the ratio
	\begin{equation}\label{eq2_4}
		R_{\text{Apsel}}=\sqrt{1-\frac{v^{2}}{c^{2}}}-\frac{Ze^{2}}{4\pi\epsilon_{0}mc^{2}r}=\left[1-(\alpha Z)^{2}\right]^{1 / 2}-(\alpha Z)^{2},
	\end{equation}
	where $v$ is the speed at which a muon undergoes uniform circular motion in a Bohr orbit. $\sqrt{1-v^{2}/c^{2}}$ represents the time dilation caused by special relativity, while $-Ze^{2}/(4\pi\epsilon_{0}mc^{2}r)$ accounts for the time delay due to the electromagnetic field. $\alpha=e^{2}/4\pi\epsilon_{0}\hbar c$ is the fine-structure constant. Table.~\ref{tab1} shows the experimental results and D. Apsel's theoretical predictions.
	
	\begin{table}[h!]
		\centering
		\caption{$R_{\text{exp}}$ is the experimental result summarized in~\cite{p10}, $R_{\text{Apsel}}$ is the calculations of Eq.~\ref{eq2_4}.}
		\setlength{\tabcolsep}{6mm}{
			\begin{tabular}{ccc}
				\hline $Z$ & $R_{\text {Apsel}}$ & $R_{\text {exp }}$ \\
				\hline $6~\mathrm{C}$ & 0.997&  1.00(2)\\
				$13~\mathrm{Al}$ & 0.986 & 0.99(4)\\
				$20~\mathrm{Ca}$ & 0.968 & 1.00(3)\\
				$26~\mathrm{Fe}$ & 0.946 & 0.97(4)\\
				$30~\mathrm{Zn}$ & 0.928 & 0.95(3)\\
				$42~\mathrm{Mo}$ & 0.858 & 0.93(5)\\
				$48~\mathrm{Cd}$ & 0.814 & 0.84(7)\\
				$53~\mathrm{I}$ & 0.772 & 0.66(7)\\
				$74~\mathrm{W}$ & 0.550 & 0.53(5)\\
				$82~\mathrm{Pb}$ & 0.443 & 0.40(10)\\
				\hline
		\end{tabular}}
		\label{tab1}
	\end{table}
	
	Now, we want to claim that the experimental results of Cd was reported as 0.90(5) or 0.84(7), here we just enumerate 0.84(7) because this number got supports by more references~\cite{p10}. D. Apsel's theory has matched the experimental results very well. However, it is well known that the Bohr model is only a semi-classical, semi-quantum theory. Moreover, adding a electromagnetic time dilation (analogous to the gravitational time dilation) is questionable. Therefore, an extension of quantum theory with a reasonable spacetime relation is in demand to resolve this problem.
	
	This article is organized as follows. In Sect.~\ref{sec2}, the basic frame is introduced. In Sect.~\ref{sec3}, we derive the Schr$\ddot{\text{o}}$dinger's equation. In Sect.~\ref{sec4}, the ratio of bound muon decay is calculated. The conclusion is given in Sect.~\ref{sec5}.
	
	\section{Basic frame}\label{sec2}
	
	Enlightened by theory of Standard Model (SM) as well as some other theories~\cite{action1,SM1,action2,action3,action4,SM2,action5,SM3,action6,action7}, considering a spin-independent particle moving in the electromagnetic field under Einstein gravity, apart from all the known actions, we construct a new field action as
	\begin{widetext}
		\begin{equation}\label{eq3_1}
			I=\int\sqrt{-g}\lambda\left[-\frac{1}{2}g^{\mu\nu}\partial_{\mu}\zeta\partial_{\nu}\zeta +\frac{1}{2}\left(\frac{mc}{\hbar}\right)^2\left( \zeta^2-\eta\right)\mathcal{D}\\ +\frac{1}{8\eta}g^{\mu\nu}\partial_{\mu}\eta\partial_{\nu}\eta\right]d^4x,
		\end{equation}
	\end{widetext}
	and the action describing the motion of a particle in electromagnetic field is
	\begin{equation}\label{eq3_2}
		S=\int Ld\tau=\int\left( \frac{1}{2}mg_{\mu\nu}\dot{x}^{\mu}\dot{x}^{\nu}+qA_{\mu}\dot{x}^{\mu}-\frac{1}{2}mc^{2}\mathcal{D}\right) d\tau .
	\end{equation}
	Where $g^{\mu\nu}$ is the inverse tensor of Riemann metric, $g$ is the determinant of $g_{\mu\nu}$, $\lambda$ is a constant with the dimension of length. $\mathcal{D}\left(x\right)$ is a scalar function that we especially proposed in this work for the spacetime relation. $\zeta$ is a scalar. $\eta$ is the proper probability density function of the particle.
	
	It is reasonable to require $\eta$ to satisfy the conservation law
	\begin{equation}\label{eq3_3}
		\partial_{\mu}\left(\sqrt{-g}\eta u^{\mu}\right)=0.
	\end{equation}
	Similar to quantum mechanics (QM), we require $\eta$ to be positive, continuous, finite and single-valued at every point of space.
	
	In the subsequent derivation we adopt natural units $\hbar=c=1$ unless otherwise stated.

	\section{Schr$\ddot{\text{O}}$dinger's equation}\label{sec3}
	
	A variation of Eq.~\ref{eq3_1} with respect to $\mathcal{D}$ leads to
	\begin{equation}\label{eq3_4}
		\zeta^2=\eta.
	\end{equation}
	Varying the action Eq.~\ref{eq3_1} with respect to $\zeta$, we get
	\begin{equation}\label{eq3_5}
		\partial_{\mu}\left(\sqrt{-g}g^{\mu\nu}\partial_{\nu}\zeta\right)+m^2\sqrt{-g}\mathcal{D}\zeta=0.
	\end{equation}
	We would like to point out that varying Eq.~\ref{eq3_1} with respect to $\eta$ only yields another form of Eq.~\ref{eq3_5}, which does not contribute any new equations.
	
	\hspace{\fill}
	
	For such a particle, with respect to Eq.~\ref{eq3_2}, the Hamiltonian is
	\begin{equation}\label{eq3_6}
		H=\pi_{\mu}\dot{x}^{\mu}-L=\frac{1}{2}m\left( g_{\mu\nu}u^{\mu}u^{\nu}+\mathcal{D}\right),
	\end{equation}
	where
	\begin{equation}\label{eq3_7}
		\pi_{\mu}=\frac{\partial L}{\partial \dot{x}^{\mu}}=mu_{\mu}+qA_{\mu}
	\end{equation}
	is the four canonical momentum vector. Then the conservation of the Hamilton Eq.~\ref{eq3_6} leads to
	\begin{equation}\label{eq3_8}
		g_{\mu\nu}u^{\mu}u^{\nu}+\mathcal{D}=const.
	\end{equation}
	When $\mathcal{D}=0$, Eq.~\ref{eq3_8} should go back to the result of conventional general relativity, that is
	\begin{equation}\label{eq3_9}
		g_{\mu\nu}u^{\mu}u^{\nu}=-1,
	\end{equation}
	thus as a spacetime relation, Eq.~\ref{eq3_8} can be re-expressed as
	\begin{equation}\label{eq3_10}
		g_{\mu\nu}u^{\mu}u^{\nu}+\mathcal{D}=-1.
	\end{equation}
	
	In order to theoretically prove the consistency between this EQM and conventional QM, the Hamilton-Jacobi method~\cite{ham} is chosen:
	\begin{equation}\label{eq3_13}
		H\left(x^{\mu},\partial_{\mu}S\right)+\frac{\partial S}{\partial \tau}=0,
	\end{equation}
	where $S$ is the Hamilton-Jacobi function and $\pi_{\mu}$ has been substituted by $\partial_{\mu}S$. With
	\begin{equation}\label{eq3_14}
		S=W\left(x^{\mu},\pi_{\mu}\right)-a_{h}\tau,
	\end{equation}
	one has
	\begin{equation}\label{eq3_15}
		H\left( x^{\mu}, \partial_{\mu}W\right)=a_{h}=-\frac{1}{2}m.
	\end{equation}
	The quantity $W$ is Hamilton characteristic function and 
	\begin{equation}\label{eq3_16}
		\pi_{\mu}=\partial_{\mu}W,
	\end{equation}
	now Eq.~\ref{eq3_10} can be rewritten as
	\begin{equation}\label{eq3_17}
		g^{\mu\nu}\left( \partial_{\mu}W-qA_{\mu}\right )\left( \partial_{\nu}W-qA_{\nu}\right)+m^2\mathcal{D} =-m^2 ,
	\end{equation}
	and Eq.~\ref{eq3_3} can also get a new form in the same way:
	\begin{equation}\label{eq3_18}
		\partial_{\mu}\left[\sqrt{-g}\zeta^2g^{\mu\nu}\left(\partial_{\nu}W-qA_{\nu}\right)\right]=0.
	\end{equation}
	The equations Eq.~\ref{eq3_17}, Eq.~\ref{eq3_18} and Eq.~\ref{eq3_5} can serve as the new fundamental formulas for demonstrating the Schr$\ddot{\text{o}}$dinger's equation.
	
	To achieve this task, one can introduce a new function $\xi$ defined as
	\begin{equation}\label{eq3_19}
		W=-i\ln\xi,
	\end{equation}
	which is then substituted in Eq.~\ref{eq3_17} and Eq.~\ref{eq3_18}, combined with Eq.~\ref{eq3_5}, resulting in the Klein-Gordon equation:
	\begin{widetext}
	\begin{equation}\label{eq3_20}
		\frac{1}{\sqrt{-g}}\partial_{\mu}\left[ \sqrt{-g}g^{\mu\nu}\left(\partial_{\nu}-iqA_{\nu}\right)\phi\right]-g^{\mu\nu}iqA_{\mu}\left( \partial_{\nu}-iqA_{\nu}\right)\phi-m^2\phi=0,
	\end{equation}
	\end{widetext}
	where 
	\begin{equation}\label{eq3_21}
		\phi=\zeta\xi=\zeta e^{iW}
	\end{equation}
	is the wave-function of the particle, and there is
	\begin{equation}\label{eq3_22}
		\eta=\zeta^{2}=\phi^{*}\phi.
	\end{equation}
	
	From now on, our calculations will adhere to the International System of Units (SI). The form of Klein-Gordon equation in Minkowski spacetime takes
	\begin{equation}\label{eq3_23}
		\left[\frac{\hbar}{i} \partial_{\mu}-q A_{\mu}\right]\left[\frac{\hbar}{i} \partial_{\mu}-q A_{\mu}\right] \phi+m^{2} c^{2} \phi=0.
	\end{equation}

	Using Eq.~\ref{eq3_4} and Eq.~\ref{eq3_5}, $\mathcal{D}\left( x\right) $ in Minkowski space can be expressed as
	\begin{equation}\label{eq3_24}
		\mathcal{D}=-\frac{\hbar^{2}}{m^{2}c^{2}}\left(\partial_{\sigma}\partial_{\sigma}\ln\sqrt{\eta}+\partial_{\sigma}\ln\sqrt{\eta}\partial_{\sigma}\ln\sqrt{\eta}\right),
	\end{equation}
	and Eq.~\ref{eq3_10} can be expressed as
	\begin{equation}\label{eq3_new_1}
		u_{\mu}u_{\mu}+\mathcal{D}c^{2}=-c^{2}.
	\end{equation}
	
	For non-relativistic approximation, substitute 
	\begin{equation}\label{eq3_26}
		\phi=\Psi e^{-\frac{i m c^{2} t}{\hbar}} 
	\end{equation}
	into Eq.~\ref{eq3_23} and ignore the higher-order infinitesimal terms, we get Schr$\ddot{\text{o}}$dinger's non-relativistic equation
	\begin{equation}\label{eq3_27}
		i \hbar \frac{\partial \Psi}{\partial t}=\frac{1}{2 m}\left[\frac{\hbar}{i} \nabla-q \vec{A}\right]^{2} \Psi+q \varphi \Psi,
	\end{equation}
	where $\Psi$ is the wave function of Schr$\ddot{\text{o}}$dinger's equation, $\vec{A}$ is magnetic vector potential and $\varphi$ is electric potential. One can see this EQM naturally results in the fundamental equation in QM. 
	
	\section{Calculation of bound muons decay ratio}\label{sec4}

 	By solving  Schr$\ddot{\text{o}}$dinger's equation, the ground state hydrogen-like wave function of the muonic atom is
	\begin{equation}\label{eq4_1}
		\Psi_{\mu^{-}}=\sqrt{\frac{1}{\pi}}\left(\frac{Z}{r_{0}}\right) ^\frac{3}{2}e^{-\frac{Zr}{r_{0}}}e^{-\frac{iE_{1}t}{\hbar}},
	\end{equation}
	where $r_{0}$ is the Bohr radius and $E_{1}$ is the ground-state energy of the muonic atom.

	By substituting Eq.~\ref{eq3_22}, Eq.~\ref{eq3_26} and Eq.~\ref{eq4_1} into Eq.~\ref{eq3_24}, we have
	\begin{equation}\label{eq4_2}
		\mathcal{D}_{\mu^{-}}=\left(\alpha Z\right)^{2}\left(\frac{2 r_{0}}{Z r}-1\right).
	\end{equation}
	In this case, Eq.~\ref{eq3_7} and Eq.~\ref{eq3_16} provide
	\begin{equation}\label{eq4_3}
		\vec{u}=\frac{\hbar}{m}\nabla W.
	\end{equation}
	By Eq.~\ref{eq3_21}, Eq.~\ref{eq3_26} and Eq.~\ref{eq4_1}, we have $\nabla W_{\mu^{-}}=0$, thus
	\begin{equation}\label{eq4_4}
		\vec{u}_{\mu^{-}}=0.
	\end{equation}
	Then the time delay can be obtained from Eq.~\ref{eq3_new_1}: 
	\begin{equation}\label{eq4_5}
		\gamma=\frac{dt}{d\tau}=\sqrt{1+\mathcal{D}}.
	\end{equation}
	Thus the average value of decay ratio $R_{\text{th}}$ could be calculated as
	\begin{align}\label{eq4_6}
		R_{\text{th}}&=\langle\frac{1}{\gamma}\rangle_{\mu^{-}}\nonumber \\
		&=\int_{0}^{\infty} \frac{1}{\sqrt{1+\mathcal{D}_{\mu^{-}}}} \Psi_{\mu^{-}}^{*}\Psi_{\mu^{-}} d^{3} x\nonumber \\
		&=\frac{Z^{3}}{\pi r_{0}^{3}} \int_{0}^{\infty} \frac{1}{\sqrt{1+\mathcal{D}_{\mu^{-}}}} e^{-\frac{2 Z r}{r_{0}}} \cdot 4 \pi r^{2} d r.
	\end{align}
	We show the results of the theoretical decay ratio changing with the atomic number $Z$ in Table~\ref{tab2}, where $R_{\text{exp}}$ is the experimental value of the muon decay ratio~\cite{p10}. It can be observed that the experimental values $R_{\text{exp}}$ decreases as the atomic number $Z$ increases, and the theoretical values are close to the experimental values when $Z$ is less than 20. However, as $Z$ increases further, the experimental values of the decay ratio become significantly smaller than the theoretically predicted values. These differences may be attributed to the increased probability of muons entering the nucleus as the atomic number increases, which increases the probability of muon absorption by protons in the nucleus~\cite{p11,p12,p17}.
	\begin{table}[h!]
		\centering
		\caption{$R_{\text{exp}}$ is the experimental result of the decay ratio, $R_{\text{th}}$ is the calculations of Eq.~\ref{eq4_6} and $R_{\text{rev}}$ is the calculations of Eq.~\ref{eq4_7}.}
		\setlength{\tabcolsep}{3mm}{
			\begin{tabular}{cccc}
				\hline $Z$ & $R_{\text {th}}$ & $R_{\text {rev}}$ & $R_{\text {exp}}$\\
				\hline $6~\mathrm{C}$ & 0.999 & 0.999 & 1.00(2)\\
				$13~\mathrm{Al}$ & 0.996 & 0.995 & 0.99(4)\\
				$20~\mathrm{Ca}$ & 0.990 & 0.986 & 1.00(3)\\
				$26~\mathrm{Fe}$ & 0.984 & 0.972 & 0.97(4)\\
				$30~\mathrm{Zn}$ & 0.979 & 0.958 & 0.95(3)\\
				$42~\mathrm{Mo}$ & 0.962 & 0.888 & 0.93(5)\\
				$48~\mathrm{Cd}$ & 0.953 & 0.838 & 0.84(7)\\
				$53~\mathrm{I}$ & 0.945 & 0.789 & 0.66(7)\\
				$74~\mathrm{W}$ & 0.911 & 0.556  & 0.53(5)\\
				$82~\mathrm{Pb}$ & 0.899 & 0.467 & 0.40(10)\\
				\hline
		\end{tabular}}
		\label{tab2}
	\end{table}
	Considering the muon entering the nucleus cannot be completely absorbed~\cite{p11,p12,p17},  the decay ratio should be revised as
	\begin{equation}\label{eq4_7}
		R_{\text{rev}}=R_{\text{th}}-\tilde{a}\int_{0}^{r_{N}}\frac{Z^{3}}{\pi r_{0}^{3}}\frac{1}{\sqrt{1+\mathcal{D}_{\mu^{-}}}} e^{-\frac{2 Z r}{r_{0}}} \cdot 4 \pi r^{2} d r
	\end{equation}
	where $r_{N}=1.2\sqrt[3]{A}\;{\rm fm}$ is the nuclear radius of the nucleus, $A$ is the mass number of the nucleus and $\tilde{a}$ is the absorption rate of muon in the nucleus.
	\begin{figure}[htbp]
		\centering
		\includegraphics[width=7cm]{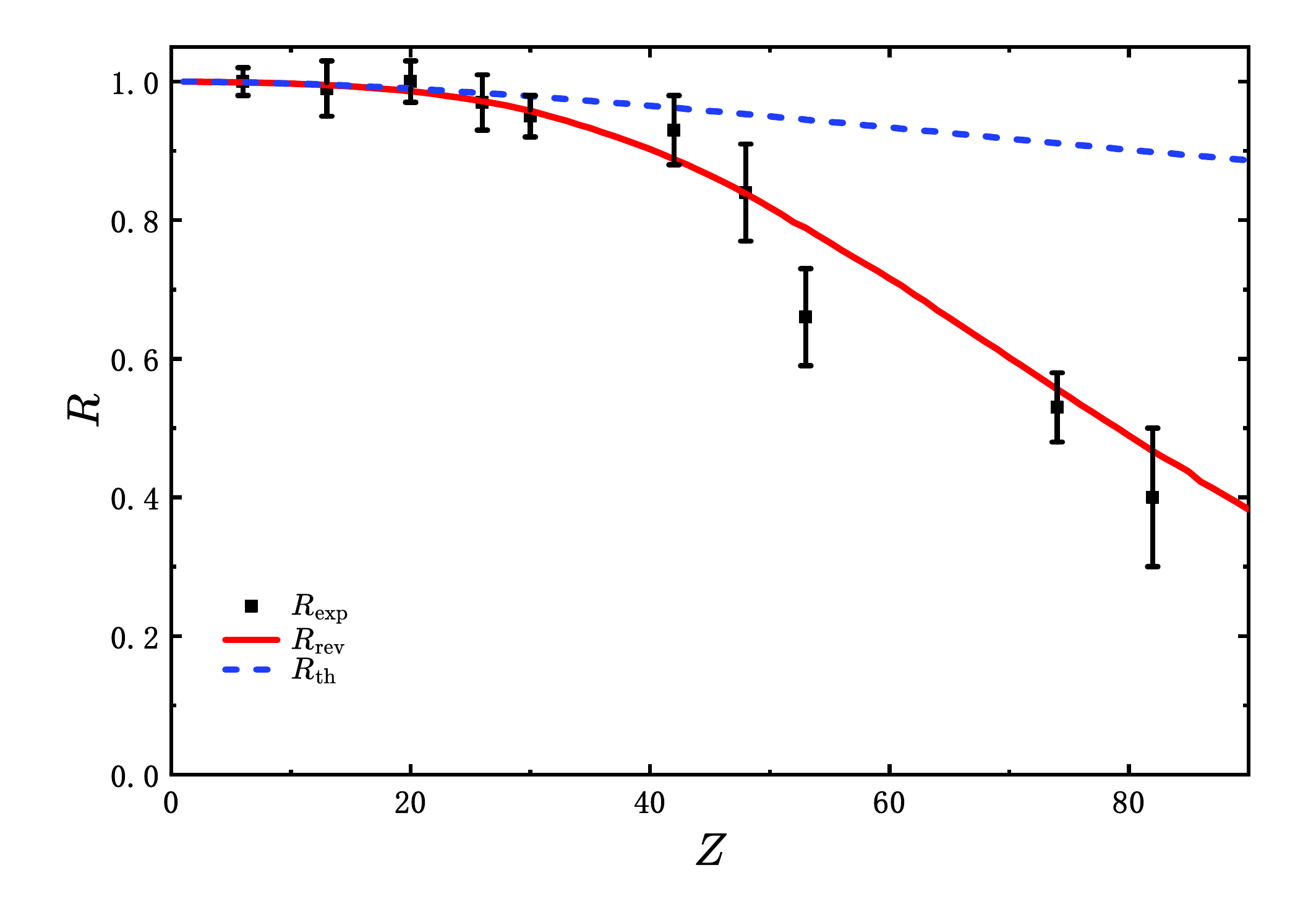}
		\caption{The ratio of the decay rate of bound muons to that of free muons, plotted against the number of binding nucleus. The spots are experimental results, the dotted line is the calculation of Eq.~\ref{eq4_6} and the solid line is the calculation of Eq.~\ref{eq4_7}.}\label{fig1}
	\end{figure}
	To the first order of our consideration, the absorption coefficient should be proportional to $Z$ because the main capture process is $\mu^- p\rightarrow\nu_{\mu} n$~\cite{p11,p12,p17}. After a lot of data fitting and careful consideration about other studies (H. {\"U}berall~\cite{p4} and D. Apsel~\cite{p10}, etc), we summarized
	\begin{equation}\label{eq4_8}
		\tilde{a}=\alpha Z,
	\end{equation}
	the corresponding results of $R_{\text{rev}}$ are listed in the third column of Table~\ref{tab1} and also plotted in Figure~\ref{fig1}. It can be seen that our modified formula is in agreement with the experimental data. 
	
	There might be more satisfactory results if one considers the effect of neutron numbers. Additionally, considering other effects such as the Coulomb effect, spin effect, etc., may also improve the accuracy of the results.
	
	\section{Conclusions}\label{sec5}
	In this paper, to resolve the decay rate of bound muons, a scheme of EQM is proposed. To prove the validation of this attempt theoretically, the Schr$\ddot{\text{o}}$dinger's equation is obtained. Then the decay ratio of bound muons is calculated, the result is in good agreement with the experimental data. 
	
	\section*{Conflicts of Interest}
	The authors declare that there are no conflicts of interest regarding the publication of this paper.
	
	\section*{Acknowledgments}
	We are grateful to Cong Li, He-Xu Zhang, Hao-Ran Zhang, Tian-Chi Ma, Lei You and Yu-Cheng Tang for useful conversations.
	
	\bibliography{paper}

\end{document}